# Approximate Queries and Representations for Large Data Sequences*

Hagit Shatkay Stanley B. Zdonik
Department of Computer Science
Brown University
Providence, RI 02912
{hs,sbz}@cs.brown.edu

**Abstract**

*Many new database application domains such as experimental sciences and medicine are characterized by large sequences as their main form of data. Using approximate representation can significantly reduce the required storage and search space. A good choice of representation, can support a broad new class of approximate queries, needed in these domains. These queries are concerned with application-dependent features of the data as opposed to the actual sampled points. We introduce a new notion of* generalized approximate queries *and a general* divide and conquer *approach that supports them. This approach uses families of real-valued functions as an approximate representation. We present an algorithm for realizing our technique, and the results of applying it to medical cardiology data.*

## 1 Introduction

Application domains such as Medicine, Music, Seismology and experimental sciences in general, all require non-traditional database support for large data sequences, such as time series. In contrast to traditional database applications, users in these domains typically search for *patterns* within the data that fit some approximate notion of what they want. They are not interested in the exact values in the time series as much as the overall shape of some subsequences.

In this paper we introduce a new framework that facilitates a broad class of approximate queries over sequences. Previous work such as [Mot88, SWZS94, WZJS94, CS94, AFS93, FRM94] regards *approximate* queries, as queries to which the answers are not exactly what was asked for. The query defines an exact result in terms of *specific values*, which is the "best" we can expect. The actual results, however, are within some measurable distance (expressed as a metric function), from the desired one. Hence, the *queries* are actually *precise*, while the *results* may be *approximate*.

Figure 1 demonstrates this notion of approximate queries for time series data. The user looks for sequences similar to the query sequence (represented by the solid curve), and the result contains all the sequences stored in the database within the boundary defined by the dashed curves. A stored sequence is an *exact match* if it is identical to the solid curve, (i.e. *distance* = 0). This form of similarity is supported by algorithms from the domain of approximate pattern matching (e.g. [CR94]).

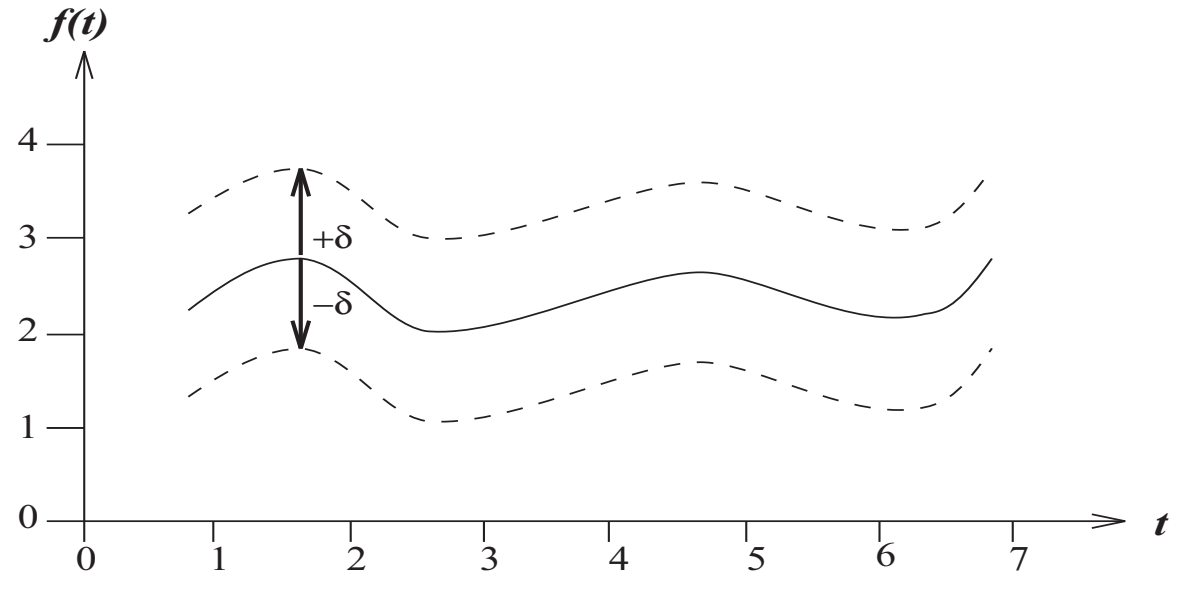

**Figure 1:** The solid curve represents a specified query sequence. The result consists of all stored sequences within distance $\delta$ from the desired sequence.

This paper presents a more general notion of approximation, appropriate for the complex queries required in advanced multimedia and scientific applications. A query mechanism in this context should allow for the specification of the general shape of data without depending on specific values. For example, in a seismic database, we may look for sudden vigorous seismic activity; in a stock market database, we look at rises and drops of stock values; in a music database, we look for a melody regardless of key and tempo. The form of approximation demonstrated in Figure 1, is value-based and does not generalize well to any of the above cases. We discuss this point in detail in Section 2.

Another aspect of these applications is the extreme length of sequences. Often, this data is archived off-line, on very slow storage media (e.g., magnetic tape), in a remote central site. Domain experts retrieve portions of the raw data, and use application programs to manipulate it. Thus, little use is made of available database management tools. For example, obtaining raw seismic data can take several days [Fis93]. The geochemist examines this data in the lab, and if it is not sufficient to answer the question at hand, another retrieval is issued.

*Partial support for this work was provided by the Advanced Research Projects Agency under contract N00014-91-J-4052 ARPA order 8225, and contract DAAB-07-91-C-Q518 under subcontract F41100.

This extremely high latency limits scientific progress.

Since the exact data points are not necessarily of interest, we can store instead an approximate representation that is much more compact, thus can be stored locally. Moreover, due to the representation's compactness, several different representations can be stored to accelerate various classes of queries.

Our method relies on the ability to break the input data into regions that can each be effectively approximated by a function (e.g., a line). From the piecewise functions we can extract features of interest (e.g., peaks). For the purpose of our example applications, drawn from the domain of medicine, polynomials proved to be sufficient. However, our approach does not rule out approximation through other well-behaved functions. We show how this technique can be highly space-efficient as well as support generalized approximation queries.

In this paper we discuss approximation both in terms of queries and in terms of representation. We generalize the previous notion of approximate queries, and suggest a strategy for handling this kind of approximation, without committing ourselves to any specific query language. Due to space limitations, we don't go into all of the details such as a full discussion of the breaking algorithms (see Section 5), or the preprocessing that needs to be performed on the raw data. Instead, we present a framework that has been tested on cardiology data using a specific representation that has shown to work well for our purposes. The same framework can be used with other approximating functions.

In Section 2 we motivate and provide a definition of generalized approximate queries. Section 3 provides a survey of related work. Section 4 presents our "divide and conquer" strategy for handling sequences, and demonstrates how we can apply it to generalized approximate queries. Section 5 presents one of the algorithms we have implemented, and provides experimental results for electrocardiograms. Section 6 provides an outline of future work.

## 2 Generalized Approximate Queries

In current application domains that use sequential data there is a need to search for patterns rather than explicit values. That is – individual *values* are usually not important but the *relationships* between them are. In what follows we present an example and analyze it to clarify the issues.

### 2.1 GoalPost Fever – an Example

One of the symptoms of Hodgkin's disease is a temperature pattern, known as "goalpost fever", that peaks exactly twice within 24 hours. A physician looking for all patients with "goalpost fever" is interested in finding all 24-hour temperature graphs that look roughly like the curve plotted in Figure 2.

Suppose we have a specific exemplar of a sequence with 2 peaks, like the one depicted in Figure 2, fixed on a cartesian system, as shown in Figure 3.

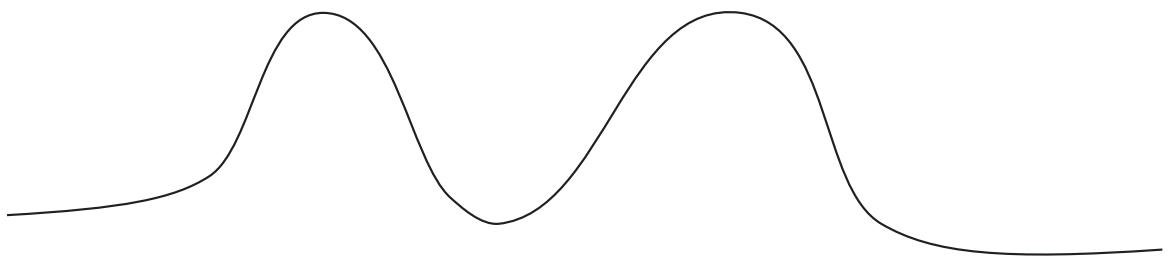

**Figure 2:** Temperature pattern with exactly two peaks.

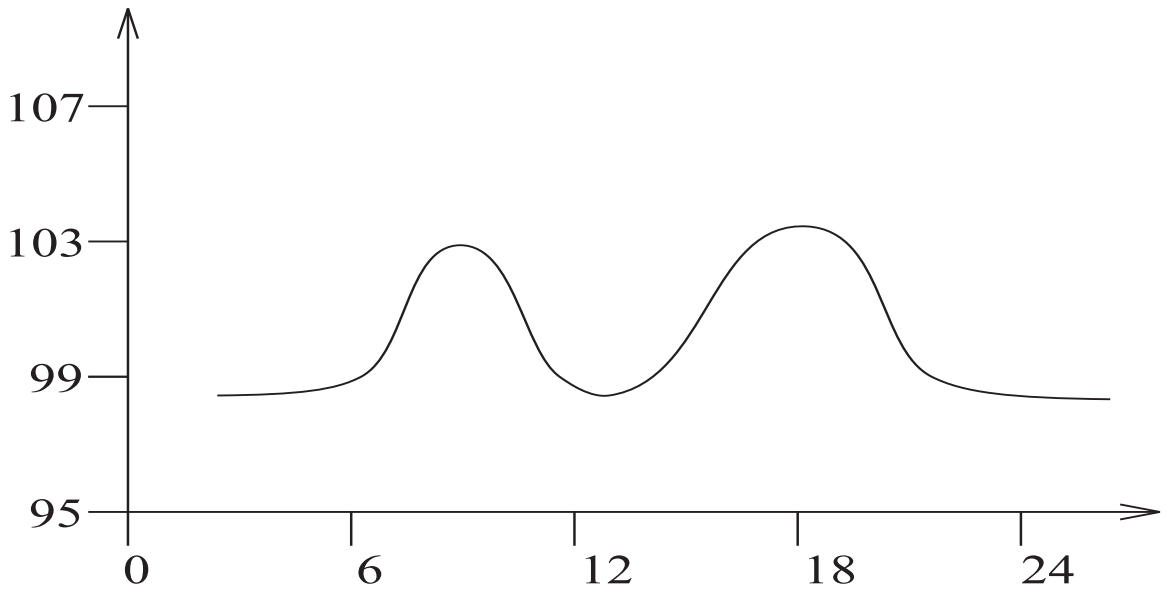

**Figure 3:** A fixed 2-peaks pattern

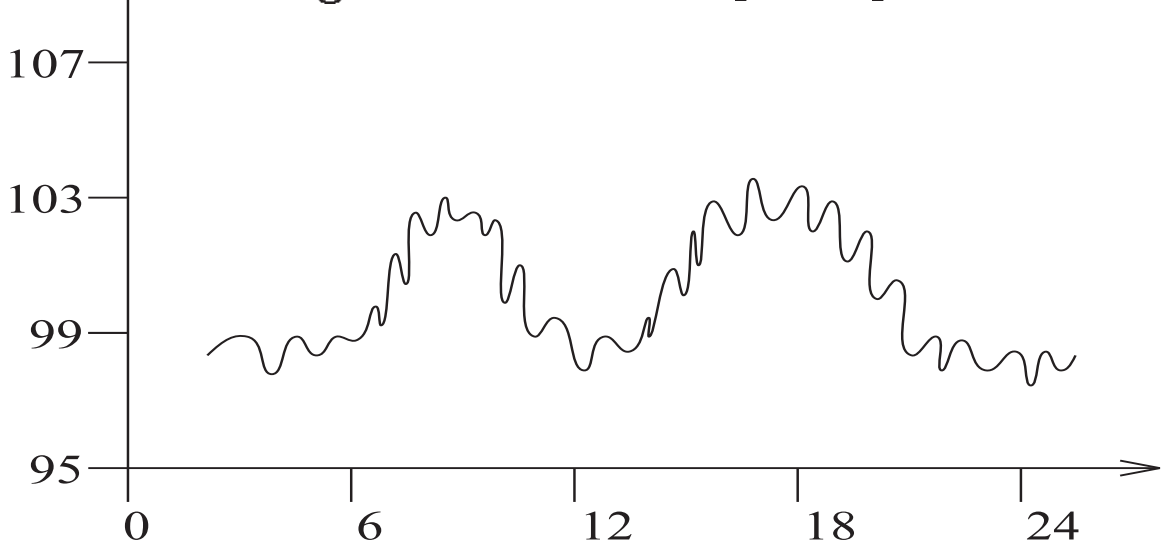

**Figure 4:** A fixed 2-peaks pattern with pointwise fluctuations within some tolerable distance.

Using this sequence as a query, fixing the distance to be $\delta$ and looking for all sequences whose values are within $\delta$ distance, allows for the matching of sequences like the one in Figure 4.

However, it does not cover sequences like those shown on Figure 5. This is because the latter demonstrate not only a $\delta$ change in amplitude with respect to the values of the query sequence, but also *scaling* (1,2,3,4),

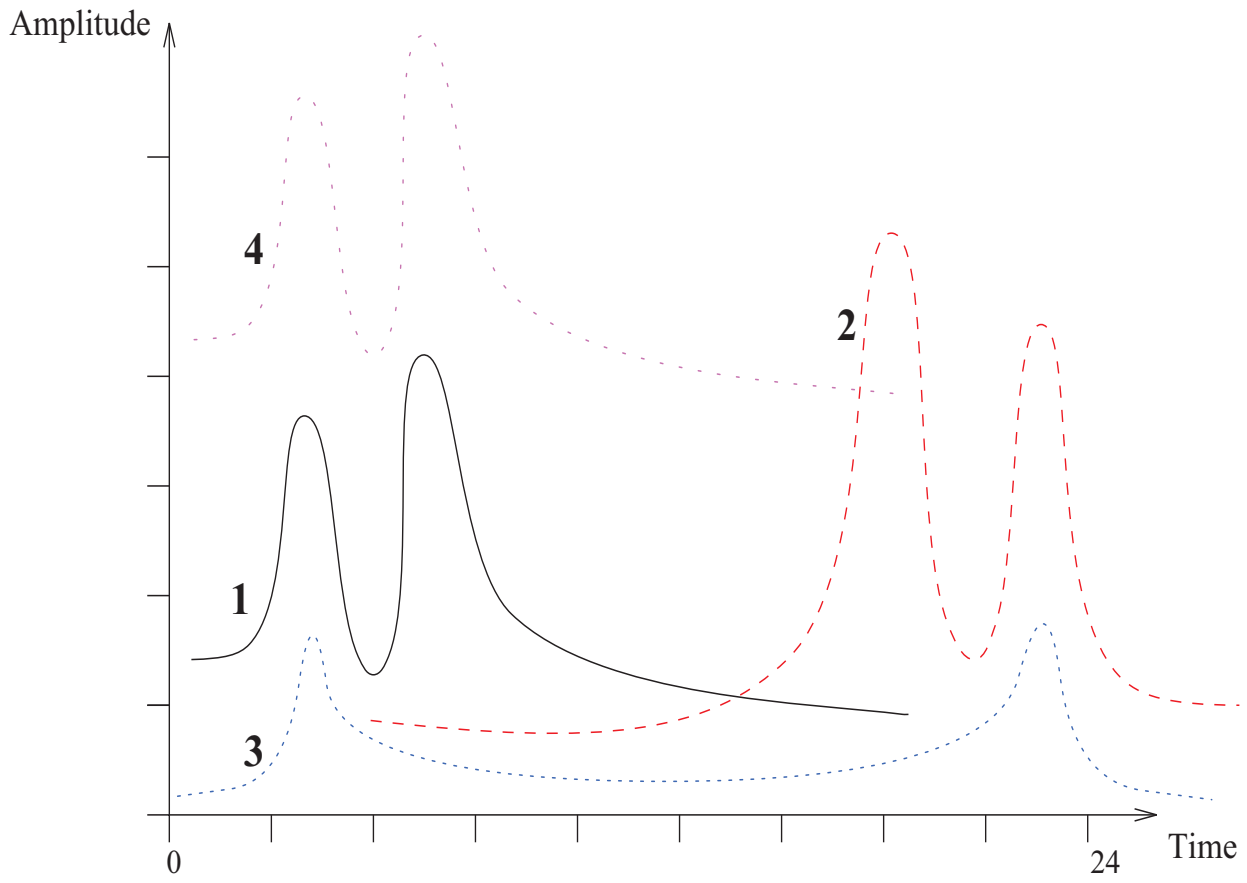

**Figure 5:** Various two-peaked sequences, *not* within a value-based distance $\delta$ from the sequence of Fig. 3

*contraction* (1,2,4), *dilation* [1](3), *shift in time* (2), *shift in amplitude* (4).

All the sequences of Figure 5 are results of *transformations* that maintain the "two peaks" property. The goalpost fever query denotes an *equivalence class* of sequences under the relation "having two peaks", which is closed under transformations that preserve this property. Since all these sequences are members of the class, they constitute an *exact match* for the query, rather than an approximation. This is despite the fact that none of them is within *value-based* distance 0 from the others.

We note that the query itself is *approximate* in the sense that it ignores specific values, and looks for "something with two peaks". The following section formalizes this notion of approximation.

### 2.2 Generalized Approximate Queries - Characterization

We define *generalized approximate queries* as queries with the following characteristics:

1. There is some general, value independent, pattern that characterizes the desired results. Defining the pattern constitutes the query. (The query can be an exemplar or an expression denoting a pattern).
2. The query denotes a (possibly uncountable) *set* $S$ of sequences, rather than a single sequence, to be matched.
3. $S$ is closed under any *behavior-preserving transformations*, and not merely under *identity* of specific values. Taking any sequence $r \in S$, any other $r' \in S$ can be obtained from $r$, through the application of transformations that preserve the significant features of $r$. The actual features and transformations are domain dependent. Examples of such transformations include:
   - Translation in time and amplitude.
   - Dilation and Contraction (frequency changes)
   - Deviations in time, amplitude and frequency.
   - Any combination of the above.
4. A result is an *exact match* if it is a member of $S$. Hence, the expected result is a *set* of matches that can all be exact.

   A result is *approximate*, if it deviates from the specified pattern, in any of the dimensions which correspond to the specified features (such as number of peaks, or the steepness of the slopes, or the minimal height required for being a peak in our example), within a domain-dependent error tolerance. The error tolerance must be a metric function defined over each dimension. This means that an approximate result is obtained from some $r \in S$ by a transformation that is not completely feature-preserving.

[1] Dilation and contraction are both forms of changes in frequency.

The above definition describes an *approximation* notion since it abstracts away from particular values and allows us to talk about how things "approximately look". It generalizes the standard notion of approximation ( [Mot88, CS94, AFS93]) in the following ways:

- Generalizes what the *query* denotes, from a single sequence (or a set closed under identity of values), to a set of similar sequences, which can be obtained from an exemplar through *similarity-preserving* transformations. Thus, we define *approximate queries* and not just *approximate results*.
- An *approximate result* can deviate from an exact one in various dimensions (each dimension corresponds to some feature), as opposed to deviation within a fixed distance from the specific values.

Moreover, it generalizes more recent notion of *similarity* over sequences (see Section 3 and [GK95, ALSS95]) from proximity under shifting/scaling of sequences, to closure under *any* form of feature preserving transformations.

### 2.3 Approximate Data Representation

Sequences such as time series are characterized by their large volume. Data is constantly generated, sampled, gathered and analyzed. The observation that data is *sampled* already hints that the actual values just "happened to be" what they are. A slight delay in the starting point of sampling can produce different values.

Moreover, efficient access is required only to subsequences of the data that satisfy domain-dependent properties, rather than to particular explicit points. Hence, it is not necessary (nor is it feasible) to store all the data in a manner that supports immediate access to any part of it at all times. It is sufficient to characterize "interesting" features in the data, and store these characterizations rather than whole sequences.

Examples of such characterizations are differential equations, approximating polynomials, compressions, or main frequencies of DFT [FRM94]. Applications may require different *alphabets* of interesting features. For example, in order to find a word in a database of spoken recorded text, it is a good strategy to transform the sequence of sampled audio signals into a sequence of words, represented as strings of ASCII characters. These strings constitute the feature alphabet.

The representation in terms of the target alphabet should have the following characteristics:

- Significantly more space efficient than the original.
- If $a, b$ are two sequences, *Features(a), Features(b)* are their respective interesting features, and *Rep(a), Rep(b)* their respective representations in the new alphabet, then: $Rep(a) = Rep(b) \iff Features(a) = Features(b)$[2]

[2] $\iff$ is a strong requirement, and can be relaxed by requiring it to hold with high probability. Probabilities assigned to each direction need not be the same. Sometimes false hits are acceptable but false dismissals are not, (thus right to left should have a higher probability), or vice versa.

Namely, similar representation should correspond to similar features.

- Preserves important features of the sequence.
- Is conveniently indexable.
- Supports queries on general patterns rather than concrete values.
- Can be used to predict/deduce unsampled points. (This is not necessary for supporting generalized approximate queries.)

We don't propose discarding the actual sequences. They can be stored archivally and used when finer resolution is needed.

## 3 Related Work

The work presented here lies in the intersection of two domains, namely, sequence support in databases and similarity (and approximation) queries. A lot of the recent work on databases for temporal and sequential data [Ric92, GS93, SS93, SZLV93, SLR94] does not address approximation. Very interesting work was done on similarity and approximation-based search for data other than sequences [Mum87, Jag91, SWZS94], but it does not generalize well to sequences.

Similarity search on sequential data is presented in [AFS93, FRM94]. This work is based on mapping *all* subsequences of fixed length to $K$-dimensional points, that are $K$ coefficients of the DFT (Discrete Fourier Transform), and using minimal bounding rectangles for storage and indexing. Queries are sequences that are matched against stored data up to some error tolerance, measured using Euclidean distance. The work is extended in [GK95] to allow for shifting and scaling of sequence amplitude. (A similar extension, using the $L_\infty$ metric, without DFT, is presented in [ALSS95]). This approach finds efficient approximate representation for time series, under the assumption that low frequencies constitute data, and high frequency is noise. However, similarity tests relying on proximity in the frequency domain, can not detect similarity under transformations such as dilation (frequency reduction) or contraction (frequency increase). For instance, looking at the goalpost fever example, *none* of the sequences of Figure 5 matches the sequence given in Figure 3, if main frequencies are compared. Moreover, their approach is based on indexing over all fixed-length subsequences of each sequence. We claim that not all subsequences are of interest, thus there is no need to facilitate efficient access to *all* subsequences.

The above work is extended in another direction in [FL95] to deduce the $K$-dimensional representation from the given data, using a provided distance function for the data, and reducing this distance to a $K$-dimensional Euclidean distance. This technique is very useful in cases where the distance function is well defined but costly to compute. However, in many cases (like in the goalpost fever case) obtaining distance functions on the actual data sequences is very hard, making the above technique not appropriate. Our approach pursues a complementary (feature based) direction, of transforming the original sequences into a simpler form from which the relevant features for comparison are easily obtained. These features are compared in order to find similarities, and distances are measured between the values of the features to accommodate approximate matches.

Other recent work presented in [ABV94] deals with recognition, matching and indexing handwritten text. They break handwritten text into letters, and incorporate Hidden Markov Models that recognize letters into a lexicographic index structure to support search. This approach doesn't generalize well to other sequential data which is not as structured and predictable as handwritten text.

Very recently Jagadish et al [JMM95] introduced a general framework, for approximate queries. It consists of a pattern language $P$, a transformation rule language $T$, and a query language $L$. Distance between objects is measured in terms of the cost of the transformation rules that map from one object to another. The framework is similar to ours in viewing similarity queries as general patterns that the data should satisfy, and in regarding transformations as a similarity measure over data sequences. However, we add a framework for representing data to support similarity queries *efficiently*. Working strictly within their framework, which uses transformation rules as an effective measurement for similarity, search is exponential at best. Our framework gives a very general similarity criterion that is not meant to be effective[3]. To compensate for this, we enhance the framework with a domain-dependent method for achieving approximate representation, tailored for the specific features that are preserved by the transformations.

Our approach is focused around prominent features, the transformations we consider are those that preserve them, and approximations are deviations from these features. In [JMM95] the transformation rules correspond to the deviations and differences between objects, rather than to "sameness". Our tailoring of the representation around the features, already embeds the predicates associated with "pattern" queries of [JMM95], into the represented data. This facilitates indexing and efficient search, and allows for the transformation-rule based approximation of [JMM95], to be treated as a simple, quantified deviation from the stored data, and therefore to be handled as regular range queries.

## 4 Our Approach – Divide and Conquer

In this section we introduce and demonstrate our idea of representing sequences in a way that facilitates generalized approximate queries. It consists of breaking sequences into meaningful subsequences and represent-ing them using well behaved real-valued functions. No other database system that we know of takes this approach.

[3] Given an arbitrary transformation $t$, and two sequences $a$, $b$, whether $t(a) = b$ is undecidable.

### 4.1 The General Approach

To facilitate the approximation discussed in Section 2, given any application domain, the following concerns must be addressed (with the help of domain experts):

- Identify key domain features of the data.
- Find a feature-preserving representation.
- Find a transformation from the time-series to the chosen representation.

Centering representation around features of interest, allows querying them directly, as demonstrated in Section 4.4. Like in any other approach for representing, storing and querying data, there is a close and limiting relationship between the stored data, and the kind of queries that are efficiently supported. Thus, the method we pursue for supporting approximation is always tightly coupled with the application domain, although it is general enough to be usefully applied in more than a single domain.

### 4.2 Function Sequences

Our technique consists of mapping digitized sequences to sequences of real-valued functions. We claim that sequences can be broken into meaningful subsequences, and each subsequence can be represented as a continuous and differentiable function. The data model underlying this approach must preserve and support sequential (temporal or other) ordering, as in [SZLV93, SS93, GS93].

Functions have the following desirable features:

1. Significant compression can be achieved. The exact compression rate depends on the nature of the data, the tolerated information loss, and the chosen functions.
2. Simple lexicographic ordering/indexing exists within a single family of functions.[4] Some examples are:

   *Polynomials* – By degrees and coefficients (where degrees are more significant)

   $$3x^2 + 2x + 1 < x^4 + 2 < 2x^4$$

   *Sinusoids* – By amplitude, frequency, phase.
3. Continuity allows interpolation of unsampled points.
4. Behavior of functions is captured by derivatives, inflection points, extrema, etc.

The last of those features is the most important for supporting generalized approximate queries. The behavior of sequences is represented through the well-understood behavior of functions.

[4] Each subsequence has to be shifted and regarded as if starting from time 0 to allow comparison of representing functions.

We associate with each application domain the specific family of functions that is best suited for it. This raises several issues such as breaking the sequences, choosing functions, storing them and indexing them. The first issue is crucial for addressing the others. The next subsection elaborates on it.

### 4.3 Breaking Up Sequences

Obtaining good representation of sequences using our technique, relies on a careful choice of the places where sequences are broken. A *breaking algorithm* determines where each subsequence starts and ends, and a *breakpoint* is a point in the sequence on which a new subsequence starts, or a previous subsequence ends. The breaking algorithm must decide to which subsequence the breakpoint belongs, according to the behavior of the resulting subsequences. The following list provides properties that a breaking algorithm must satisfy to be beneficial:

**Consistent** – Sequences with similar features (where 'similarity' is domain-dependent) are broken at corresponding breakpoints, with respect to those features.

Thus, the features by which subsequences are determined (for instance – minima and maxima points) are detected even when the sequence is modified by any form of feature-preserving transformations. Therefore, if $A$ and $B$ are two sequences such that $A$ can be obtained from $B$ through a feature-preserving transformation, then when $A$ and $B$ are broken into subsequences, each subsequence of $A$ can be obtained through a transformation of the corresponding subsequence of $B$.

**Robust** – Minor changes to a sequence do not affect the respective breakpoints[5]. Putting it formally:

Let $S$ denote the sequence $< s_1, ..., s_n >$. Let $S'$ denote the sequence obtained from $S$ by adding a new element $s'$ between $s_l$, $s_{l+1}$ $(0 \leq l \leq n)$[6].

1. Suppose $< s_i, ..., s_{l-1}, s_l, s_{l+1}, ..., s_j >$ is one of the subsequences obtained from applying the breaking algorithm to $S$, and $F(t)$ is the representing function (within error tolerance $\epsilon$) of $< s_i, ..., s_j >$. If there exists $t$, $l < t < l+1$, s.t. $F(t) - s' < \epsilon$, then the algorithm must break $S'$ s.t. $< s_i, ..., s_{l-1}, s_l, s', s_{l+1}, ..., s_j >$ is one of the subsequences. The rest of the subsequences are the same as those of $S$.
2. Suppose $< s_i, ..., s_l >$ and $< s_{l+1}, ..., s_j >$ are two consecutive subsequences of $S$, and $F_1(t), F_2(t)$ are their respective representing functions. If there exists $t$, $l < t < l+1$,

[5] To achieve robustness various kinds of preprocessing are applied to the sequences prior to breaking, such as filtering for eliminating noise, normalizing and compression . A full discussion of the methods and results is beyond the scope of this paper.

[6] If $l = 0$, $s'$ is added right before $s_1$. If $l = n$, $s'$ is added right after $s_n$.

s.t. $F_1(t) - s' < \epsilon$, (or $F_2(t) - s' < \epsilon$), then the algorithm must break $S'$ s.t. $< s_i, ..., s_l, s' >$ (or $< s', s_{l+1}, ..., s_j >$ respectively) is one of the subsequences. The rest of the subsequences are the same as those of $S$. (If the condition holds for *both* $F_1$ and $F_2$, any of the two associations of $s'$ preserves robustness).

Therefore, adding or deleting "behavior preserving" elements to the sequence, where the behavior is captured by the representing function, does no more than shift the breakpoints by at most the number of elements added/deleted.

**Avoids Fragmentation** – Most resulting subsequences should be of length >> 1. This is necessary for achieving a substantial compression.

We have implemented and experimented with several breaking algorithms, which are demonstrated and discussed in Sections 4.4 and 5.

### 4.4 GoalPost Fever – an Example (cont.)

To illustrate how the divide and conquer approach handles generalized approximate queries, we show how it is applied to the Goalpost fever query, discussed in Section 2.1. The query looks for all 24 hours sequences with *exactly two peaks*. To keep things simple we assume the following:

1. Each original sequence of 24 hour temperature logs is broken at extremum points, where little local extrema are ignored – up to error tolerance $\epsilon$. Since *peaks* are features of interest in the medical domain, it is a very reasonable strategy.

   One of the algorithms we developed and implemented uses linear interpolation to break sequences (see Section 5), and satisfies the above assumption – as demonstrated in Figure 6.

2. The resulting subsequences are represented by linear approximations. This is easily achieved by either using the interpolation line – a by-product of the breaking algorithm, or by calculating the linear regression line through each subsequence. (Other information included in the representation, like start/end points of subsequences is of no interest for this example). Figure 6 shows the results of running our breaking program on a sequence, where the program calculates the approximating regression line of each resulting subsequence.

3. An index structure that supports pattern matching (like the ones discussed in [Fre60, AHU74, Sub95]) is maintained on the "positiveness" of the functions' slopes. For a fixed small number $\phi$, there are 3 possible index values: $+\phi$ (slope $> \phi$), $-\phi$ (slope $< -\phi$), or 0 (slope is between $-\phi$ and $\phi$). We take $\phi = 0.3$. For example, given the sequence $0(+\phi)0(-\phi)0(+\phi)(-\phi)$ (over the alphabet $\{+\phi, -\phi, 0\}$), by using the index, we get the positions of the first point of all stored sequences that match that pattern. (Other index structures are not required for this example).

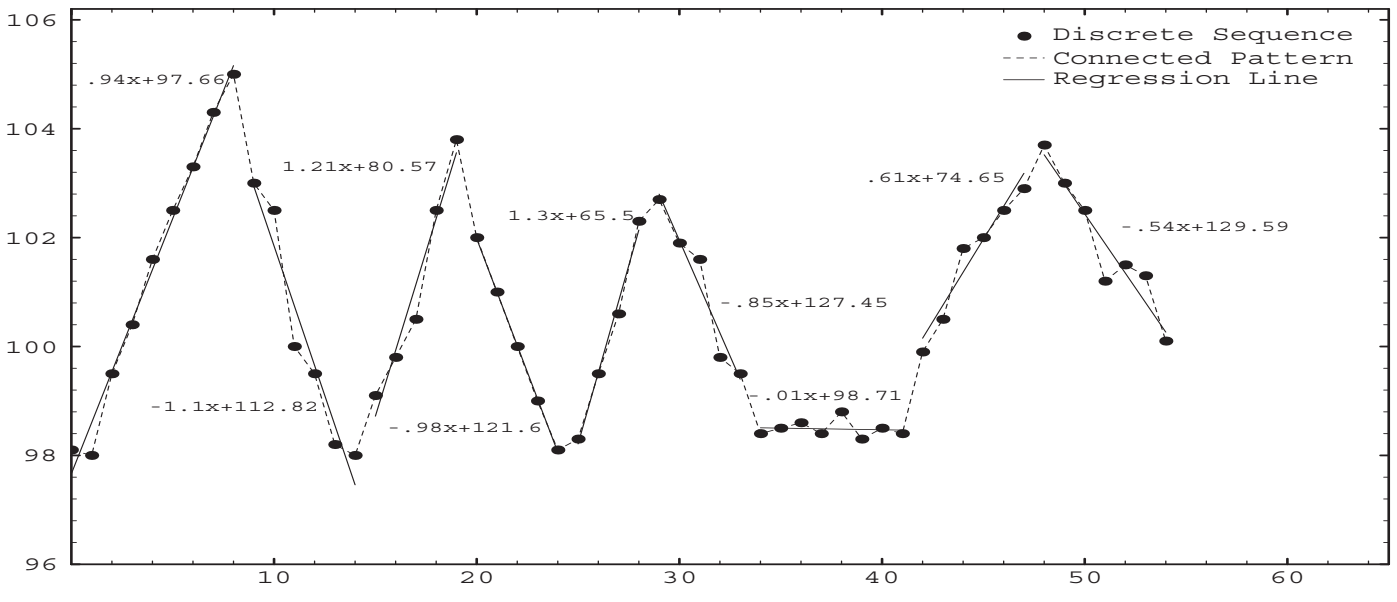

**Figure 6:** Breaking a sequence at extrema and representing it by regression functions. The function is specified near each line.

Our approach works as follows:

**The database** – The stored sequences are represented as *sequences of linear functions*. Each function is an approximation of a subsequence of the original sequence, as demonstrated in Figure 6. It is important to note that although for the particular query of finding peaks it is sufficient to store only the positions of peaks, or even just the number of existing peaks (as opposed to complete functions), this would have left us with too little information to answer any other useful queries regarding the behavior of the temperature between the peaks, or to approximately reconstruct the original data.

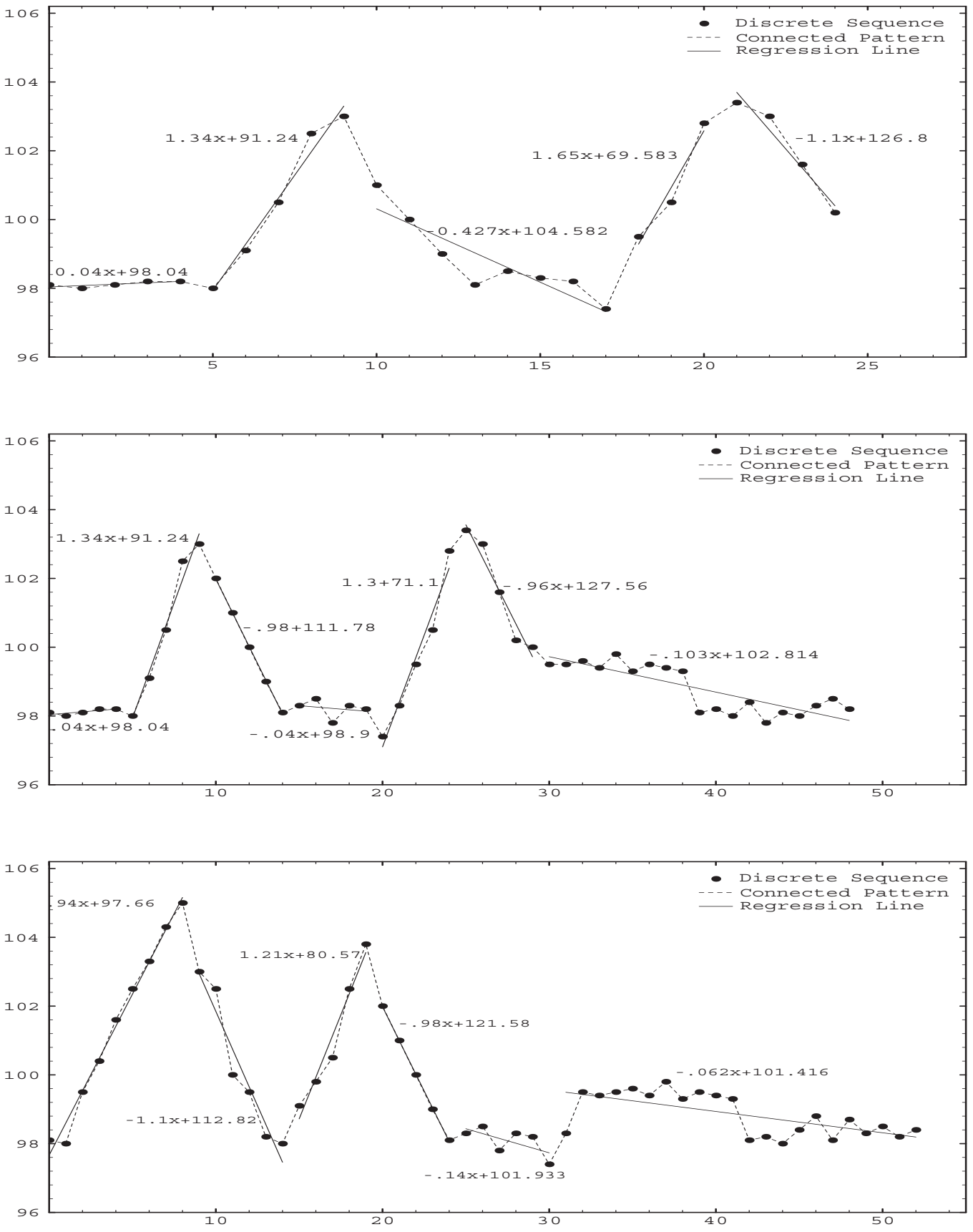

**Figure 7:** Three two-peaks sequences, broken at extrema by our algorithm, and approximated by regression lines.

**The Query** – The most naive way to pose the query is as a regular expression over the alphabet $\{+\phi, -\phi, 0\}$, as defined previously:

$$0^*(+\phi)(-\phi)0^*(+\phi)(-\phi)0^*$$

We point out that our approach does not depend on this particular choice of pattern language.

**The Result** – Following the index structure, we can find all sequences matching the required pattern. The sequence depicted in Figure 6, does not match the query pattern, while those depicted in Figure 7 are all exact matches.

Note that the correctness of the results depends on $\phi$ (the steepness of the slopes) and the distance tolerated between the linear approximation and the subsequences. Our program was parameterized with distance 2.00 for the above example[7]. These values are application dependent.

Any peak-preserving transformations (such as the ones listed in Section 2.1) applied to any of the sequences of Figures 6 and 7, results in similar breakpoints and linear representation of the respective sequences, (proof is based on the breaking algorithm), and therefore the results for the above query would be only those sequences with exactly two (prominent) peaks.

# 5 Breaking Algorithms and Results

The goal of the breaking algorithms is to obtain breakpoints in places where the behavior of the sequence changes significantly. The first part of this section surveys the algorithms we use, while the second shows the results of applying one of them to digitized electrocardiograms (ECGs).

## 5.1 Algorithms Review

There are two classes of algorithms we have studied and implemented:

**On-line algorithms**, determine breakpoints while data is being gathered, based on the data seen so far, with no overall view of the sequence [Kae94]. Their main merit is that an additional step of post-processing is not required. Their obvious deficiency is possible lack of accuracy. Hence, it is difficult to come up with an on-line algorithm that satisfies all our requirements for a wide variety of sequences. We implemented and studied one family of on-line algorithms based on sliding a window, interpolating a polynomial through it, and breaking the sequence whenever it deviates significantly from the polynomial. Our experiments are documented in [Sha95], and we are still studying algorithms using a related approach.

**Off-line algorithms**, are applied to complete sequences. The basic template of the algorithms we use is given in Figure 8. It is a generalization of an algorithm for Bézier curve fitting [Sch90]. Any type of curve, $c$ (such as polynomials of a fixed degree – rather than the Bézier curves in the original algorithm), can be used within it, resulting in subsequences of the original sequence $S$, each of which can be approximated by a curve of type $c$. In addition to being restricted to Bézier

```
Curve Fitting Algorithm

Let c be a type of curve.
begin
    Global Variables:
    S   Sequence of points ((x_1, y_1), ..., (x_n, y_n))
    ε   Error tolerance
      1. Fit a curve of type c to S
      2. Find point (x_i, y_i) in S with maximum
         deviation from curve.
      3. If deviation < ε return(S).
      4. Else {
         (a) Fit a curve c to the subsequence
             ending at (x_{i-1}, y_{i-1}), S'.
         (b) Fit a curve c to the subsequence
             starting at (x_{i+1}, y_{i+1}), S''
         (c) If (x_i, y_i) is closer to the curve ob-
             tained in (a)
             make (x_i, y_i) the last element of S'
             Else make (x_i, y_i) the first element
             of S''.
         (d) Recursively apply the algorithm to
             S' and S''.
         }
end
```

**Figure 8**: The general template of the curve fitting algorithm

curves, (requiring a parameterization step prior to step 1 of Figure 8), the original algorithm imposed continuity between curves, thus associating the breakpoint found in step 2 of Figure 8 with both resulting subsequences. Our application doesn't require continuity, and we want to prevent the breakpoint from appearing as both the end of one subsequence and the beginning of the next. Steps 4(a)-(c) are another adjustment we made to the original algorithm, to decide with which subsequence to associate the breakpoint.

We have instantiated the curve type, ($c$ of Figure 8), in three ways – a modified Bézier curve[8], a linear-regression line, and an interpolation line through the endpoints of the respective sequences. A full report of our experiments can be found in [Sha95]. Here we concentrate on the linear interpolation algorithm, and only briefly review the *Bézier Curves* and linear regression algorithms.

*Bézier Curves* are used in computer graphics for rep-resenting digitized curves [FvDFH90]. Computer-graphics techniques match our interest in queries based on "the way sequences look" (as demonstrated in Section 2.1). They also generalize well to sequences other than time-series (not *functions* of time), and to multi-dimensional sequences. Moreover, having a graphics-oriented representation of sequences, allows the use of methods from computer graphics (such as the multi-resolution analysis) for further analysis of data. However, unlike computer graphics applications, we have no indication of where curves start/end (no "mouse clicks"), nor do we allow user interference in the breaking process. The algorithm in [Sch90] supports fully automated curve fitting of digitized curves and therefore is useful for our purposes. Its strengths and weaknesses for breaking sequences are discussed in [Sha95].

[7] This is the $\epsilon$ in Figure 8 of the next section.

[8] Implementation for Schneider's original algorithm was available through ftp wuarchive.wustl.edu at graphics/graphics/books/graphics-gems/Gems.

A simpler version of curve fitting is the use of linear functions for curves. We have experimented with both linear regression and linear interpolation of endpoints. The latter is simpler and produces better results, and is described in the remainder of this section.

The *linear interpolation* algorithm takes as its curve the line interpolating the endpoints of the sequence, and effectively breaks sequences at extremum points. The intuitive explanation is that by passing a non-vertical line through the sequence, extremum points are further away from it than others. Thus the point most distant from the line is either some maximum point above it or a minimum point below it. Due to the recursion step (Figure 8, step 4(d)), the extremum point becomes one of the endpoints in the next iteration, thus points close to it would be close to the interpolation line, and therefore – there is no fragmentation, unless it is justified by extremely abrupt changes in the sequence's value. Hence the algorithm is *robust, consistent and avoids fragmentation.* Another advantage of the algorithm is that finding an interpolation line through two points does not require complicated processing of the whole sequence. Only end points need to be considered in order to generate the line. The algorithm's run time is *O(number of peaks·n)* (where $n$ is the sequence length). It is much faster than another approach we have taken, using dynamic programming, minimizing a cost function of the form *a·(# of segments) + b·(distance from approximating line)* which runs in time $O(n^3)$.

## 5.2 Linear Interpolation on ECGs

We already demonstrated the applicability of our linear interpolation program for breaking sequences in the context of the goalpost fever query, on data we generated ourselves. We also tested our program on actual digitized segments of electrocardiograms[9] and the results are demonstrated in Figure 9. The prominent peaks in the figure are pointed to by $R$. Such breaking is useful for addressing actual queries of the form "Find all ECG's with $R$-$R$ intervals of length $n \pm \delta$", where $\delta$ is the error tolerance on the distance. (That is, distance $n \pm \delta$ between two consecutive peaks). It can be answered easily, if we have an index, such as an inverted index structure ( [Car75, FJ91] or some variation on it) based on the time elapsed between peaks. Building such an index is reasonable, since this kind of query is often required [Eli95], and can be done as follows:

1. Find the peaks in the sequences. This can be done while storing the data as part of the preprocessing – by examining the slopes of the representing functions. If data is already stored, an (existing) index structure (see [Sub95] for instance), that finds all subsequences of the form $(+\phi)(-\phi)$ (as defined in Section 4.4) and returns their positions, can be used.
2. Start and end points of subsequences are part of the information obtained from the breaking algorithm, and are maintained with any representation of the sequence. Hence, a table like Table 1 can be constructed for each sequence, in which peaks are found. The table contains for each peak the approximating functions with the positive and negative slopes, and the start and end points of the respective subsequences approximated by those functions. Each point is a pair of the form $(time, amplitude)$.
3. For each peak, compare the amplitude at the end point of the rising subsequence (REnd) with that of the start point of the descending subsequence (DStart). The one with the larger *amplitude* is where the peak actually occurred. Keep *time* for that point.

[9]The segments of ECGs are available through WWW, http://avnode.wustl.edu.

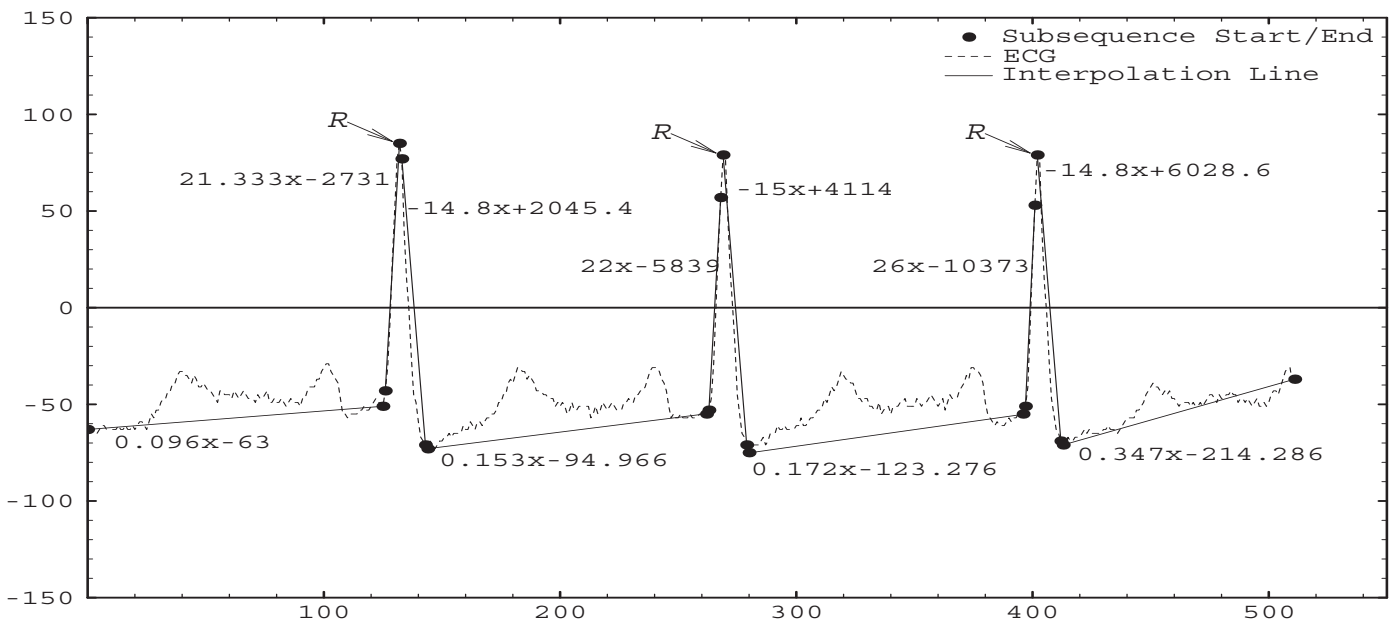

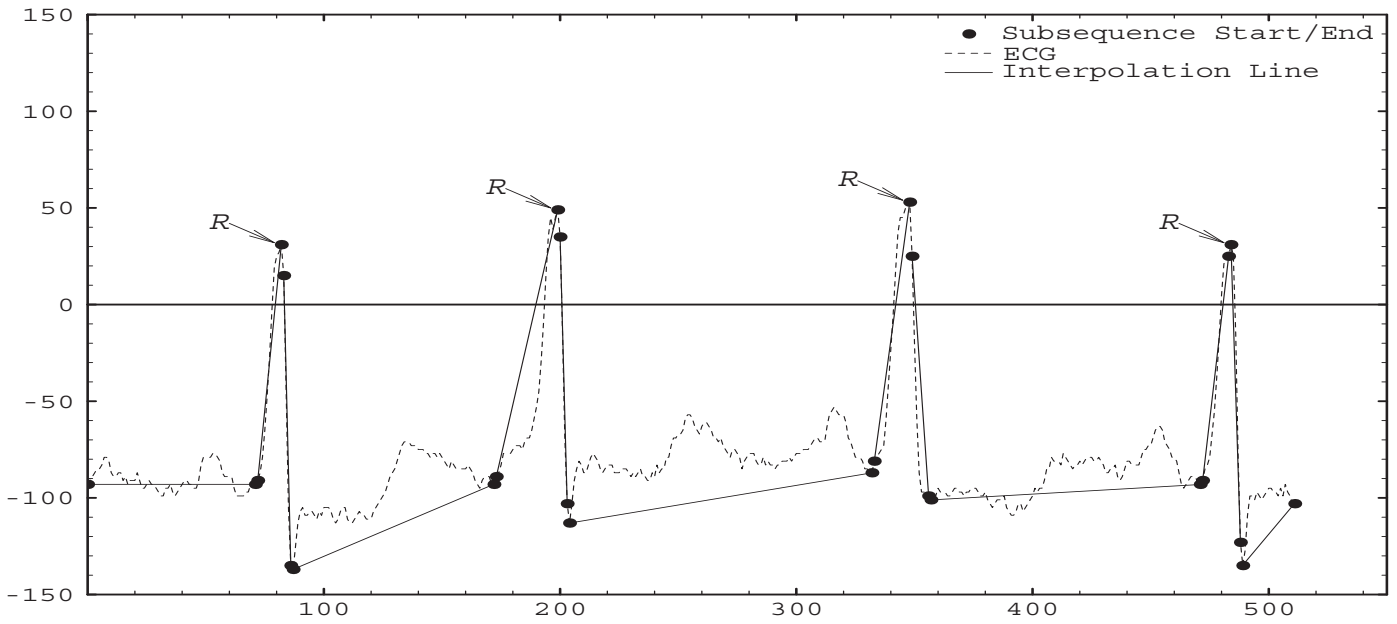

**Figure 9:** Two ECG segments of 512 points each, broken by our algorithm. The distance parameter ($\epsilon$ of Figure 8) was set to 60.

| Peak | Rising Function | RStart | REnd | Descending Function | DStart | DEnd |
|---|---|---|---|---|---|---|
| 1 | 21.333x-2731 | (126,-43) | (132,85) | -14.8x+2045.4 | (133,77) | (143, -71) |
| 2 | 22x-5839 | (263,-53) | (268,57) | -15x+4114 | (269,79) | (279, -71) |
| 3 | 26x-10373 | (397,-51) | (401,53) | -14.8x+6028.6 | (402,79) | (412, -69) |

**Table 1**: Peaks information for the top ECG on Figure 9.

4. For each pair of successive peaks, find the difference in time between them. The result is a sequence of distances between peaks. For the top ECG of figure 9, the sequence is: $< 137, 133 >$
while for the bottom one, the obtained sequence is: $< 117, 149, 136 >$.
Since the *R-R* intervals correspond to the time elapsed between every two heartbeats, the interval can not exceed a certain integer and can not go below some threshold (for any living patient). Hence there is a limited number of interval values, according to which the sequences can be indexed. A simple inverted file index is sufficient for this purpose and is used for this example. Using a more elaborate structure (see [Sub95]) would support more complex queries.
The inverted-file index structure for our data is as shown in Figure 10. It consists of a B-Tree structure, which points to the postings file. The postings file contains "buckets" of R-R interval lengths and a set of pointers to the ECG representations which contain those interval lengths. This set can also be augmented with the position of the interval of this length, but this isn't necessary, because the physician is most likely to look at the complete ECG image and see the required phenomenon on his/her own.
In order to find the ECG's with an *R-R* interval of duration $n \pm \delta$ (suppose $n = 150, \delta = 2$), we follow the B-Tree structure looking for values between 148-152, and get to the posting list for 149 (Each bucket in the postings file is sorted by the values stored in it), and find that ECG2 (the bottom ECG depicted in figure 9) satisfies the query.

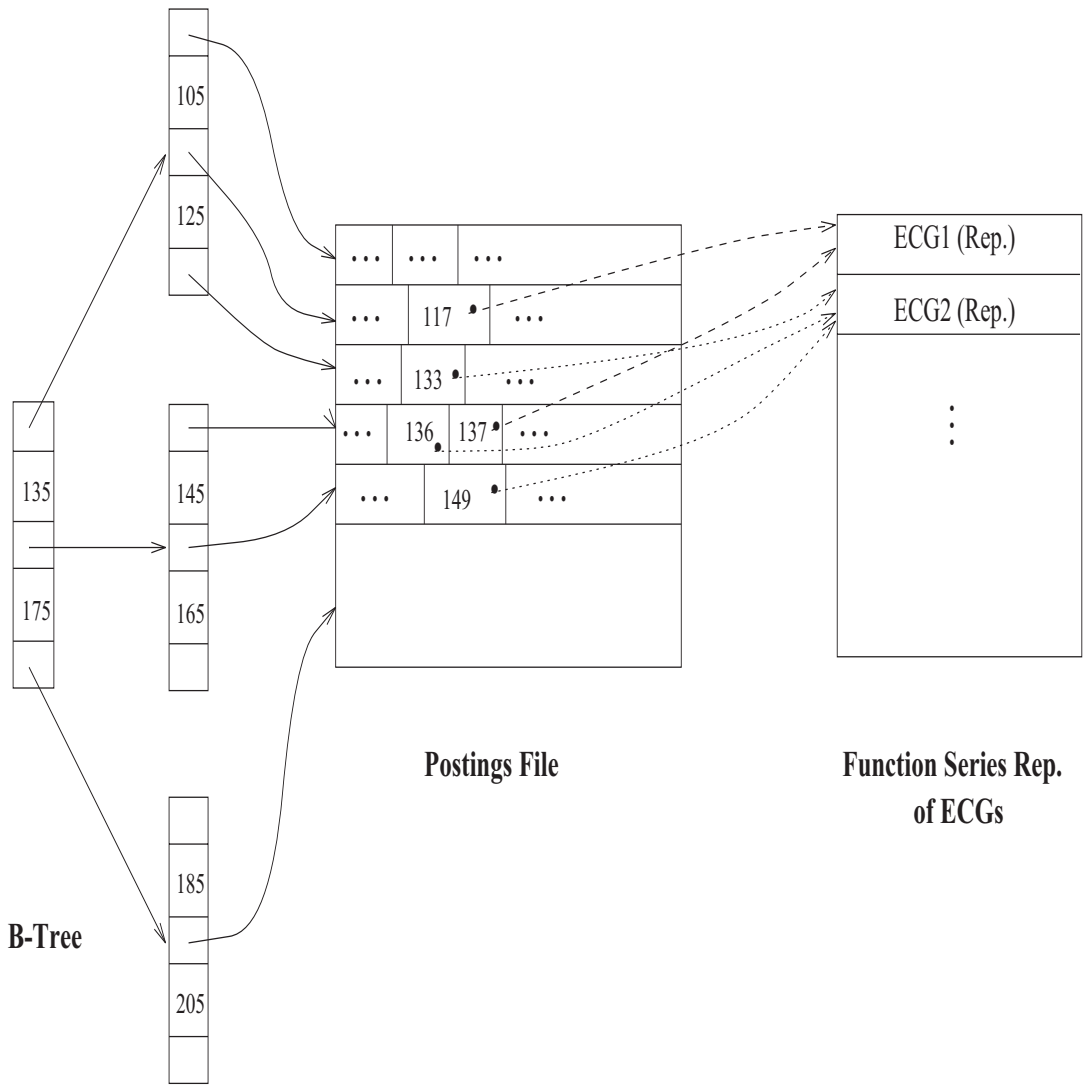

**Figure 10**: An inverted file structure for R-R intervals

The "*R-R* interval" query is a *generalized approximate query* according to the criteria set in Section 2.2:

1. The pattern characterizing the desired result is: "having distance exactly $n$ between peaks".
2. The query defines a set $S$ of all electrocardiograms having the property, regardless of explicit values.
3. The set is closed under transformations that preserve the distance between peaks, such as shift in time or in amplitude of the whole sequence.
4. A result is an *exact match* if the distance between its peaks is *exactly* $n$.
5. A result is an *approximate match* if the distance between its peaks is within $\delta$ distance from $n$. If we allow the slopes for what is considered a peak to be within some error around $\phi$ then we get another dimension of approximation – with respect to "peakness".

Figure 9 illustrates the efficiency of representation that is obtained by our technique. 512 points sequences are represented by about 12 function segments. Assuming each representation requires 4-6 parameters (such as function coefficients and breakpoints), we get about a factor of 10 reduction in space.

Thus, the example demonstrates the ability of our technique to facilitate queries unsupported within any other framework, to reduce significantly the size of a stored sequence, and to support search by using sequence-based index structures.

Our breaking algorithms produce as a by-product functions that approximate the subsequences. In some cases, the function is a good representation of the subsequence, while in other cases some other representation is required. For instance, in the example given in the Section 4.4, the by-product functions were interpolation lines, but the ones used for representation were regression lines. It is possible that with some adaptation (see Section 6) by-product functions can be a good representation.

It should be pointed out that since our representation is quite compact, it would be possible to compute and store multiple representations and indices for the same data. This would be useful for simultaneously supporting several common query forms.

## 6 Conclusions and Future Work

We have presented a new notion of approximation, which is required in application domains new to the database world, such as medicine, experimental sciences, music and many others. We address two basic needs of these domains. The need for queries based on patterns of behavior rather than on specific values, and the need to reduce the amount of stored and scanned data while increasing the speed of access.

We formally introduce the new notion of *generalized approximate queries*, and propose a framework that supports them. Our "divide and conquer" approach, is based on breaking sequences into meaningful subsequences and storing an approximate, compact representation of each subsequence as a mathematical function. Since a function's behavior can be captured by properties of the derivatives, indexing can be done according to such properties. Queries that specify sequences with certain behavior can be transformed in the same way, and can be matched to the appropriate sequences using such indices.

We presented several algorithms for breaking sequences, and demonstrated the applicability of our approach for solving real problems in medical applications. Our method also reduces the amount of data to be scanned for answering such queries.

Due to space limitations, we omitted the details of the algorithms, that we implemented for both breaking sequences and preprocessing them prior to breaking. We use various algorithms for filtering, compression (using the wavelet-transform [FS94, HJS94, Dau92]), and normalization (to have mean 0 and variance 1). Such preprocessing is useful for reducing the amount of data and for ensuring that our breaking algorithms work properly. Normalization is important both for maintaining robustness of our breaking algorithms (see Section 4), and also for enhancing similarity and eliminating the differences between sequences that are linear transformations (scaling and translation) of each other.

Future work includes:

- Continue applying our approach to additional problem domains.
- Experiment more with compressing the sequences before and after breaking while preserving their important features. Most compression techniques (such as the Lempel-Ziv algorithm) are concerned with losslessly reducing the amount of data, but not with making the compressed data have the same features (peaks for instance) as the original. Currently we are experimenting with multiresolution analysis and applying the wavelet transform, for compressing the sequences in a way that allows extracting features from the compressed data, rather than from the original sequences.
- Define a query language that supports generalized approximate queries. One of the options is to use a visual query language in which the user draws the shape of the sequence he/she is looking for, points out the important dimensions for comparison, and specifies error tolerance in each dimension. Constraint logic programming [KKR90, BJM93], or the language presented in [APWZ95], may provide a reasonable basis for an underlying query language.
- Address efficiency considerations for storage and access of data stored in the approximated format.

## Acknowledgments

We thank John Hughes for introducing us to Bézier curves and other graphics techniques, Leslie Kaelbling for pointing us in the direction of on-line breaking algorithms, Jon Elion (M.D.) from the Miriam Hospital, for the background on medicine related data problems and for the short tutorial in cardiology, Karen Fischer and John Weeks from the Geochemistry Department for sharing the data problems of seismology with us, and Catriel Beeri, Swarup Acharya, Vasiliki Chatzi, Dimitris Michailidis, Christopher Raphael, Yaron Reshef and Bharathi Subramanian for their help and advice.